\documentclass[aps,prl,floatfix,superscriptaddress,reprint]{revtex4-1} 
\usepackage{graphicx}
\usepackage{amsmath,amsfonts,amssymb}
\usepackage{color}
\usepackage{pdfpages}

\makeatletter
\AtBeginDocument{\let\LS@rot\@undefined}
\makeatother

\begin{document}

\title{On the saw-tooth torque in anisotropic $j_{\rm eff} = 1/2$ magnets: Application to $\alpha$-RuCl$_3$}
\date{\today}
\begin{abstract}
The so-called ``Kitaev candidate'' materials based on $4d^5$ and $5d^5$ metals
have recently emerged as magnetic systems displaying strongly anisotropic
exchange interactions reminiscent of the Kitaev's honeycomb model.
Recently, these materials have been shown to
commonly display a distinct saw-tooth
angular dependence of the magnetic torque over a wide range of magnetic fields.
While higher order chiral spin interactions have been considered as a source of
this observation, we show here that bilinear anisotropic interactions and/or
$g$-anisotropy are each sufficient to explain the observed torque response, which may be distinguished on the basis of high-field measurements.
These findings unify the understanding of magnetic torque experiments in a
variety of Kitaev candidate materials. 
\end{abstract}

\author{Kira Riedl}
\email{riedl@itp.uni-frankfurt.de}
\affiliation{Institut f\"ur Theoretische Physik, Goethe-Universit\"at Frankfurt,
Max-von-Laue-Strasse 1, 60438 Frankfurt am Main, Germany}
\author{Ying Li}
\affiliation{Institut f\"ur Theoretische Physik, Goethe-Universit\"at Frankfurt,
Max-von-Laue-Strasse 1, 60438 Frankfurt am Main, Germany}
\author{Stephen M. Winter}
\affiliation{Institut f\"ur Theoretische Physik, Goethe-Universit\"at Frankfurt,
Max-von-Laue-Strasse 1, 60438 Frankfurt am Main, Germany}
\author{Roser Valent{\'\i}}
\affiliation{Institut f\"ur Theoretische Physik, Goethe-Universit\"at Frankfurt,
Max-von-Laue-Strasse 1, 60438 Frankfurt am Main, Germany}

\maketitle

\begin{figure}
\includegraphics[width=\columnwidth]{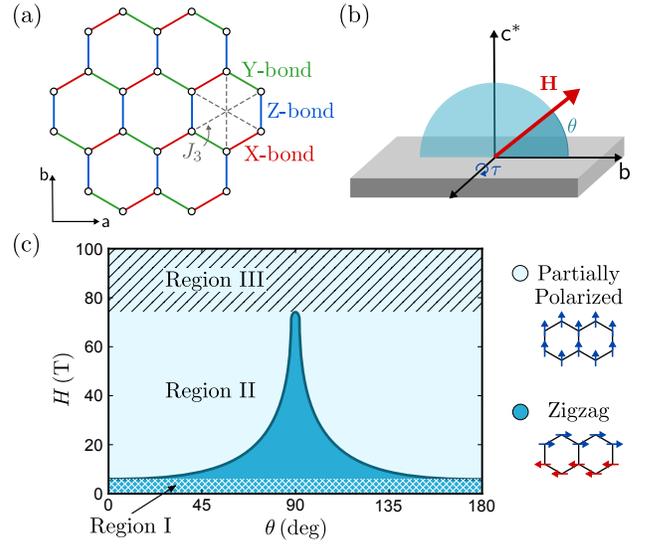}
\caption{(a) 24-site honeycomb lattice cluster used for the ED calculations with definition of the anisotropic bonds and $J_3$ interaction. 
(b) Definition of the angle $\theta$ between the field $\mathbf{H}$ and the $b$-axis in the $bc^\ast$-plane used in the torque calculations. (c) Phase diagram for the model given by Eq.~\eqref{eq:Hamil} with the parameters $(J_1,K_1,\Gamma_1,J_3)=(-0.5,-5.0,+2.5,+0.5)\,$meV and $(g_{a},g_{b},g_{c^\ast})=(2.3,2.3,1.3)$. 
Region I indicates $H<H_c(0^\circ)$, Region II indicates $H_c(0^\circ)<H<H_c(90^\circ)$, and Region III indicates $H>H_c(90^\circ)$.
} \label{fig:phase}
\end{figure}

Recently, there has been renewed interest in the study of heavy transition metal insulators with $d^5$ electronic configurations, as a possible platform for studying unique frustrated and anisotropic magnetic interactions\cite{jackeli2009mott, schaffer2016recent, rau2016spin, winter2017models,hermamms2018physics}. This interest has stemmed from possible connections to Kitaev's exactly solvable honeycomb model\cite{kitaev2006anyons}, which implies a spin-liquid ground state for various tricoordinate lattices with bond-dependent Ising interactions $\mathcal{H} = \sum_{\langle ij \rangle} S_i^\gamma S_j^\gamma$. Here $\gamma = \{x,y,z\}$ for the three bonds emerging from each lattice site (Fig.~\ref{fig:phase}(a)).
 It has been proposed that these specific interactions can arise in $d^5$ systems with local $j_{\rm eff} = 1/2$ moments\cite{jackeli2009mott,chaloupka2013zigzag,rau2014generic,rau2014trigonal} due to a delicate balance of spin-orbit coupling (SOC), Hund's coupling, and crystal-field splitting (CFS). To date, various materials have been found that realise the requisite electronic and lattice structure, such as $\alpha$-RuCl$_3$\cite{plumb2014alpha,kim2015kitaev, banerjee2016proximate, banerjee2017neutron} and the alkali iridates A$_2$IrO$_3$ (A = Na, Li)\cite{singh2010antiferromagnetic, singh2012relevance,modic2014realization,takayama2015hyperhoneycomb}. However, the presence of additional anisotropic interactions\cite{kim2016crystal,kim2015kitaev,yadav2016kitaev,winter2016challenges,hou2017unveiling} beyond the Kitaev coupling induce a variety of magnetic orders\cite{liu2011long, ye2012direct, biffin2014unconventional, biffin2014noncoplanar, johnson2015monoclinic,sears2015magnetic, williams2016incommensurate} in these systems at zero magnetic field, spoiling the potential spin-liquid. For this reason, the effects of magnetic field have been explored as a possible route towards inducing novel phases\cite{yadav2016kitaev,janssen2016honeycomb,baek2017evidence,zheng2017gapless,wolter2017field, banerjee2018excitations, ruiz2017correlated}, as magnetic order is suppressed.

In this context, the magnetic torque response provides a sensitive probe of both the phase diagram and anisotropic couplings. In general, a finite torque $\tau$ reflects an angular variation of the free energy $\mathcal{F}$ in the presence of an external field $\mathbf{H}$:
\begin{align} \label{eq:torque_gen}
\tau(\theta) = \frac{\text{d} \mathcal{F}}{\text{d} \theta}
\end{align}
where $\theta$ is the angle between $\mathbf{H}$ and a reference axis. Magnetic torque has been measured for various Kitaev-candidate materials including the 2D honeycomb  $\alpha$-RuCl$_3$\cite{leahy2017anomalous,modic2018chiral,modic2018resonant}, and Na$_2$IrO$_3$\cite{das2018magnetic}, as well as the related 3D systems $\beta$-Li$_2$IrO$_3$\cite{takayama2015hyperhoneycomb} and $\gamma$-Li$_2$IrO$_3$\cite{modic2017robust,modic2018chiral,modic2014realization}. In these materials, experiments show a distinct saw-tooth angle dependence of $\tau(\theta)$ at high fields, in addition to a number of anomalies at intermediate fields that mark field-induced phase transitions with strongly anisotropic critical fields $H_c(\theta)$. 

 \begin{figure*}[]
\includegraphics[width=\textwidth]{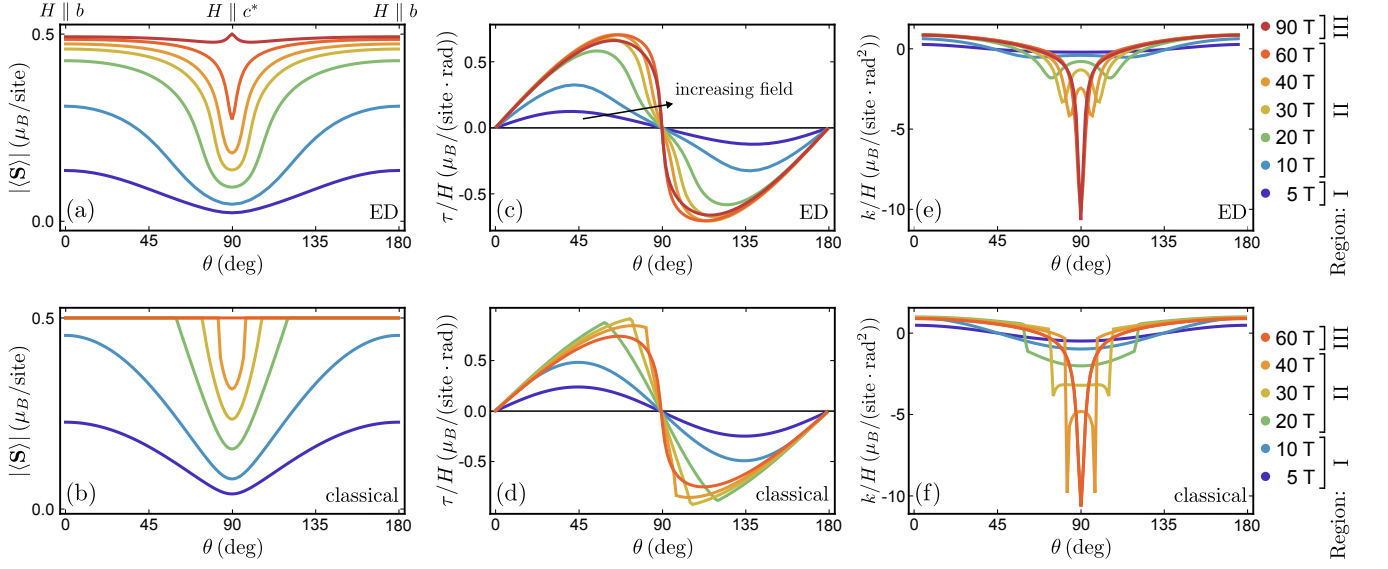}
\caption{Computed properties of the model Eq.~(\ref{eq:Hamil}) for the parameters $(J_1,K_1,\Gamma_1,J_3)=(-0.5,-5.0,+2.5,+0.5)\,$meV and $(g_{a},g_{b},g_{c^\ast})=(2.3,2.3,1.3)$.  Magnetization for several field strengths from (a) exact diagonalization  and (b) classical calculations. Torque, normalized with field for several field strengths in the $bc^\ast$-plane from (c) ED and (d) classical calculations. Magnetotropic coefficient $k=\text{d}\tau/\text{d}\theta$, normalized with field for several field strengths from (e) ED and (f) classical calculations.} \label{fig:torque_angle}
\end{figure*}

In this work, we focus mainly on $\alpha$-RuCl$_3$, for which a variety of interesting intermediate-field phases have also been proposed
to lie at selected angles\cite{liu2018dirac}, or in narrow field ranges ($\Delta H \sim 1$ T) near the edge of the low-field zigzag
order\cite{janssen2016honeycomb,banerjee2018excitations,lampen-kelley2018field}. In particular, recent thermal transport measurements\cite{kasahara2018majorana} have been interpreted in terms of a spin-liquid with Majorana edge-states at intermediate fields. Here, we consider torque as a key probe of such field-induced transitions, address experimental signatures of potential intermediate field phases, and discuss how torque may distinguish between different microscopic anisotropic interactions, more generally.

In the case of $\alpha$-RuCl$_3$\cite{leahy2017anomalous,modic2018chiral,modic2018resonant} anomalies in $\tau(\theta)$ appear for $|\mathbf{H}|\gtrsim7$ T, as the field is rotated out of the honeycomb
$ab$-plane (Fig.~\ref{fig:phase}(b)). The minimal magnetic Hamiltonian for this material includes nearest neighbour Heisenberg $J_1$, Kitaev $K_1$, and off-diagonal $\Gamma_1$ interactions, as well as longer range third neighbour $J_3$ coupling \cite{kim2015kitaev,kim2016crystal,yadav2016kitaev,winter2016challenges,winter2017breakdown,hou2017unveiling}:
\begin{align} \label{eq:Hamil}
\mathcal{H} =& \  \sum_{\langle ij \rangle} J_1 \ \mathbf{S}_i \cdot \mathbf{S}_j + K_1 S_i^\gamma S_j^\gamma + \Gamma_1 \left( S_i^\alpha S_j^\beta + S_i^\beta S_j^\alpha\right) \nonumber \\
& \ + \sum_{\langle \langle \langle ij \rangle \rangle \rangle} J_3 \ \mathbf{S}_i \cdot \mathbf{S}_j -\mu_B \sum_{i} \mathbf{H}\cdot \mathbb{G} \cdot \mathbf{S}_i.
\end{align}
where e.g. $\{\alpha,\beta,\gamma\} = \{x,y,z\}$ for the Z-bond of Fig.~\ref{fig:phase}(a). Previously, Das {\it et~al.}\cite{das2018magnetic} studied the parameter dependence of the $\tau$ for similar models, with a focus on Na$_2$IrO$_3$. For the parameters $(J_1,K_1,\Gamma_1,J_3)=(-0.5,-5.0,+2.5,+0.5)\,$meV and the anisotropic $g$-tensor $\mathbb{G}$ with $(g_a,g_b,g_{c^\ast})=(2.3,2.3,1.3)$, this model reproduces several aspects of the inelastic neutron scattering (INS)  at low temperature and zero field\cite{winter2017breakdown}, as well as INS and electron spin resonance (ESR) at finite field\cite{winter2018probing} in $\alpha$-RuCl$_3$. A similar parameterization is also consistent with the high temperature thermal hall conductivity\cite{cookmeyer2018spin}. 
It is worth noting that this minimal model Eq.~\eqref{eq:Hamil} assumes $C_3$ rotational symmetry, i.e.~all bonds have equal interaction strength. In contrast, {\it ab-initio} studies\cite{kim2016crystal,yadav2016kitaev,winter2016challenges} and various experiments (ESR\cite{little2017antiferromagnetic, wu2018field}, torque\cite{leahy2017anomalous}, susceptibility\cite{lampen-kelley2018field}) have suggested an inequivalence of the X-, Y-, and Z-bonds. In this work, we neglect this inequivalence for simplicity, as it does not modify our conclusions regarding the torque for the out-of-plane rotation depicted in Fig.~\ref{fig:phase}(b).

We find that the combination of anisotropic $g$-tensor and finite $\Gamma_1>0$ lead to a strongly anisotropic critical field $H_c(\theta)$ separating low-field antiferromagnetic zigzag order from an asymptotically polarized phase at high field, as shown in the phase diagram in Fig.~\ref{fig:phase}(c). Here, $H_c$ was estimated from extrema of $\partial^2 E/\partial H^2$ and $\partial^2 E/\partial \theta^2$ via exact diagonalization (ED) calculations on the 24-site cluster shown in Fig.~\ref{fig:phase}(a). 
For an in-plane field ($\theta=0^\circ$), the model enters the field-polarized phase at $H_c(0^\circ)\sim 6\,$T. For the out-of-plane direction ($\theta=90^\circ$) zigzag order survives up to a much higher $H_c(90^\circ)\sim 75\,$T. For the classical model, these values are modified to 11 T and 51 T, respectively. As discussed below, it is thus useful to divide the phase diagram into three regions, depending on the value of $|\mathbf{H}|$ relative to $H_c(0^\circ)$ and $H_c(90^\circ)$ as shown in Fig.~\ref{fig:phase}(c). In Fig.~\ref{fig:torque_angle} we show the angle dependence of the magnetization, torque, and ``magnetotropic coefficient'' $k = \text{d}\tau/\text{d}\theta = \text{d}^2\mathcal{F}/\text{d}\theta^2$ obtained from ED and classical calculations for different $|\mathbf{H}|$ covering these regions.  Further characterization and field dependence of $\tau$ and $k$ are presented in \cite{sup}. 

For low fields $|\mathbf{H}|< H_c(0^\circ)$ (Region I from Fig.~1(c)), antiferromagnetic zigzag order is found for all field orientations. For the chosen field rotation, the  zigzag domain with ordering wavector parallel to the monoclinic $b$-axis is uniquely stabilized. For such a domain, the local moments of the antiferromagnetic order lie in the $ac^*$-plane due to the significant $\Gamma_1>0$ in the model\cite{chaloupka2016magnetic,sizyuk2016selection}. As a result, the susceptibility is largest for fields along the $b$-axis ($\theta=0^\circ$), which is orthogonal to the zero-field ordered moments. This tendency is also reinforced by the significant $g$-anisotropy with $g_{ab}> g_{c^*}$. As a result, in Region I, the magnetization follows a smooth angle dependence with a soft minimum for the out-of-plane direction. The torque shows nearly conventional sinusoidal behaviour $\tau/H \propto (\chi_{bb}-\chi_{c^*c^*})H\sin 2\theta$ for both the ED and the classical calculations, while $k$ has no distinct anomalies. This behaviour is consistent with the experimental observations\cite{leahy2017anomalous,modic2018resonant,modic2018chiral}, and follows from the fact that the overall magnetization is small, and the uniform susceptibility tensor $\chi_{\mu\nu}$ is roughly field-independent at low field.

For intermediate field
strengths $H_c(0^\circ)< H < H_c(90^\circ)$ (Region II), spontaneous magnetic order is suppressed for some field
orientations. Strong drops in the magnetization in the vicinity of $\theta=90^\circ$ mark the phase boundary between the zigzag and polarized states. The simple $\sin 2\theta$ dependence of $\tau(\theta)$ is invalidated since angle sweeps cross this phase transition. In the classical results, phase boundaries are marked by a kink in $\tau(\theta)$, as shown in Fig.~\ref{fig:torque_angle}(d) for $H$ = 20 $-$ 40 T. These kinks give rise to jumps in the magnetotropic coefficient $k = \text{d}\tau / \text{d}\theta$, shown in Fig.~\ref{fig:torque_angle}(f). In the ED results of Figs.~\ref{fig:torque_angle}(c),(e), the kinks and jumps are rounded due to finite size effects, but appear consistent with the classical results. As discussed by Modic \textit{et al.}\cite{modic2018resonant}, phase boundaries with anisotropic $H_c(\theta)$ must be marked by a finite jump in $k(\theta)$. For example, for second order transitions, $\Delta k$ can be related to the specific heat: $\Delta k \propto - \Delta C \left[ \left( \frac{\partial T_c}{\partial H} \right)_\theta \left( \frac{\partial H_c}{\partial \theta} \right)_T \right]^2$, where $T_c$ is the critical temperature. This identifies low temperature magnetic torque as a {\it key thermodynamic probe} for detecting the presence of a distinct spin-liquid phase lying between the zigzag and polarized states $\alpha$-RuCl$_3$, as suggested by recent transport measurements\cite{kasahara2018majorana}. Such a phase is absent in the present model, so we find only a single peak in $k(\theta)$ for $0^\circ < \theta < 90^\circ$, consistent with the experimental $k(\theta)$ measurements \cite{modic2018resonant}.

For still higher fields $H > H_c(90^\circ)$ (Region III), the peaks in $k(\theta)$ merge as the zigzag order is completely suppressed for all orientations. In the polarized phase, the $c^*$ axis represents a hard axis as a result of both the finite $\Gamma_1>0$ and $g$-anisotropy. The classical magnetization is constant (polarized), while the ED result shows small dips due to enhanced fluctuations for low-symmetry field directions. The computed torque follows the saw-tooth shape observed experimentally at high field, with large (but not divergent) $k$ remaining near $\theta=90^\circ$. The sharp angle dependence of $\tau$ for nearly out-of-plane fields reflects a rapid rotation of the magnetization vector as the field passes through this hard axis. These observations demonstrate that the saw-tooth torque arises naturally when considering the anisotropic exchange and $g$-tensor terms known to be relevant to $\alpha$-RuCl$_3$ (and related Kitaev materials).

 \begin{figure}
\includegraphics[width=\columnwidth]{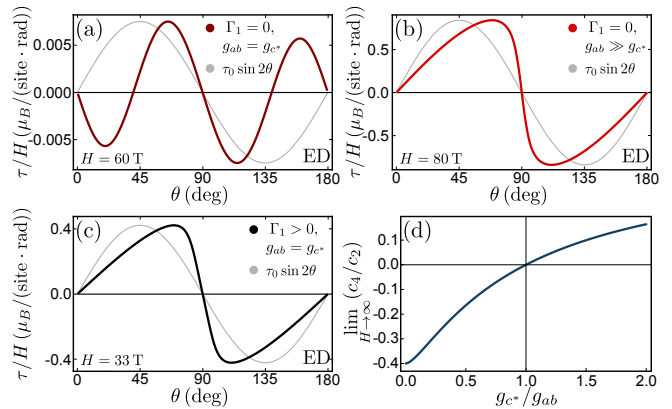}
\caption{(a)-(c): Exact diagonalization results for the magnetic torque for different models, with $\sin 2\theta$ shown in gray for comparison. For (b) and (c), $H$ corresponds to the out-of-plane critical field $H_c(90^\circ)$ of the model. (d) Dependence of the high-field limit of $c_4/c_2$ (see Eq.~\eqref{eq:torque_expansion}) on the $g$-anisotropy, which allows for extraction of $g$-values from $\tau$ data.} \label{fig:torque_models}
\end{figure}

In order to compare with experiments, it is further useful to disentangle the effects of $g$-anisotropy and different exchange anisotropies (i.e. $\Gamma_1$ and $K_1$). In general, the magnetic torque may be expanded as:
\begin{align} \label{eq:torque_expansion}
\frac{\tau}{H} = \sum_{n=1}^{\infty} c_{2n} \sin(2n\theta) = c_2 \sin(2\theta) + c_4 \sin(4\theta) + ...
\end{align}
By applying classical and ED simulations at zero temperature ($T = 0$) to different models, together with high temperature expansions, we find the scaling of constants $c_{2n}$ with $H$ and $T$ provides unique signatures for different sources of anisotropy.

We first consider a model with pure Kitaev $K_1$ anisotropy, taking $(J_1,K_1,\Gamma_1,J_3) = (-0.5,-5.6,$ $0,+0.5)$ meV and $g_a = g_{c^*}=2$. 
At the classical level, the ferromagnetic interaction $K_1<0$ cannot select any preferred magnetization axis in either the zigzag or polarized phases. As a result, the classical $\tau(\theta)$ vanishes. However, a weak cubic anisotropy emerges in the quantum model via a quantum order-by-disorder mechanism\cite{sizyuk2016selection}, which provides notable $ \sin 4\theta$ components to the out-of-plane torque, as shown in ED calculations in Fig.~\ref{fig:torque_models}(a). This mechanism also leads to anomalous scaling of the coefficients $|c_{2n}|$: all odd $n$ coefficients must scale identically to the even $n+1$ coefficients. For example, high temperature expansion suggests that both $|c_2|$ and $|c_4|$ scale as $\propto H^3/T^5$ in the limit $T \gg T_c,\frac{\mu_BH}{k_B}$. Similarly, in the low-temperature polarized Region III, we find from ED calculations that $|c_2|, |c_4| \propto H^{-2}$, and $|c_6|, |c_8|\propto H^{-4}$. While these weak cubic anisotropies are overpowered in known ``Kitaev'' materials by other interactions, it is noteworthy that quantum order-by-disorder effects can generally be diagnosed from anomalous scaling of $|c_{2n}|$.

 \begin{figure}[]
\includegraphics[width=\linewidth]{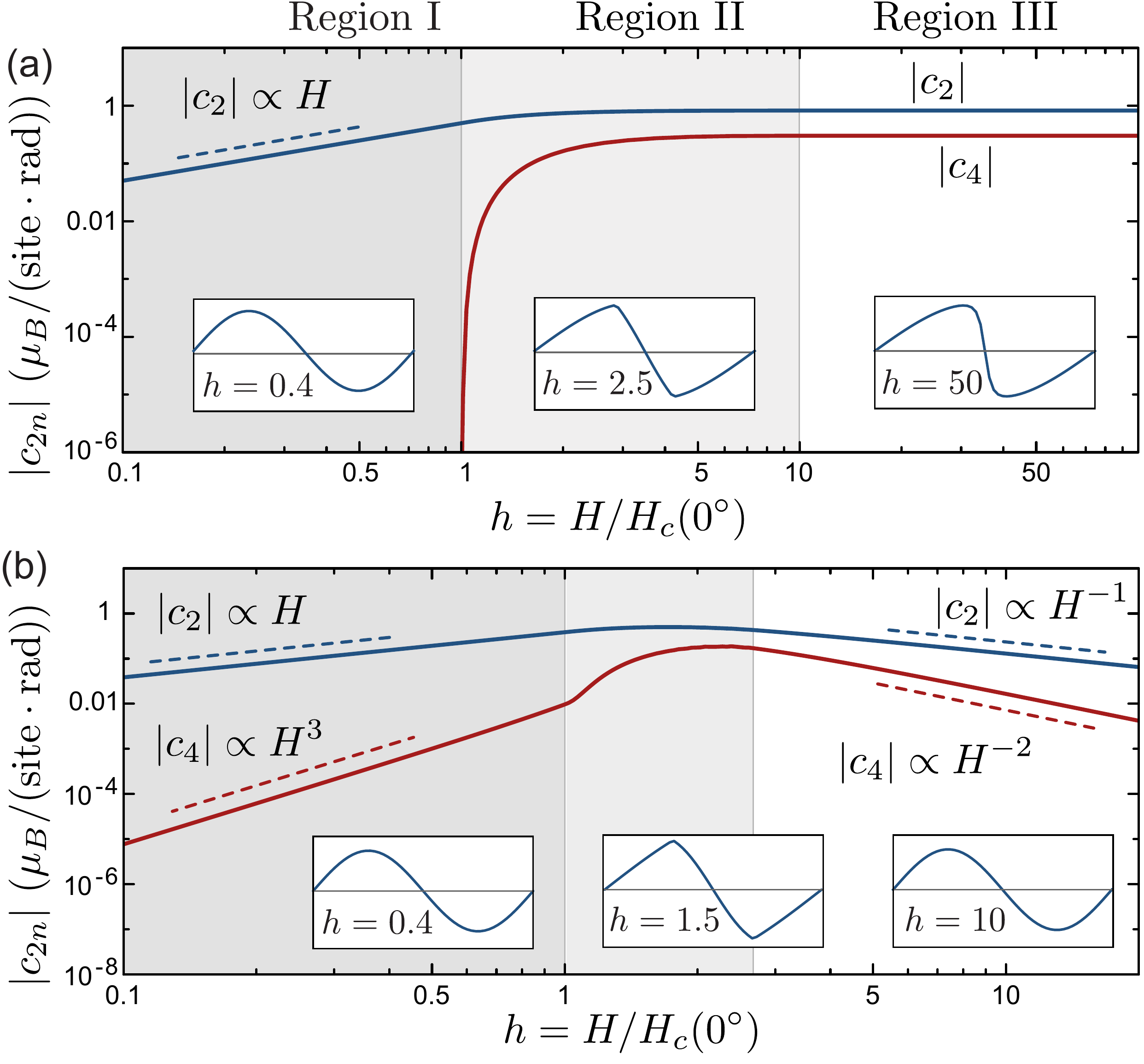}
\caption{Classical field dependence of coefficients $|c_{2n}|$ for models where torque arises from (a) pure $g$-anisotropy, and (b) pure bilinear $\Gamma_1$ exchange. Insets show example $\tau(\theta)$ curves in Regions I - III, as defined in Fig.~\ref{fig:phase}(c).} \label{fig:torque_scale}
\end{figure}

Next, we consider the same model as above, but with $g_a = 2, g_{c^*} = 0.2$. In this case, the $g$-anisotropy dominates, leading to a significant saw-tooth form of $\tau(\theta)$, as shown in Fig.~\ref{fig:torque_models}(b) for ED calculations at $H = H_c(90^\circ)$. In Fig.~\ref{fig:torque_scale}(a), we show the {\it classical} scaling of $|c_{2n}|$. Since $K_1$ does not contribute to $\tau$ at the classical level, the classical torque arises only from the $g$-anisotropy in this model. In the low-field ordered phase (Region I with $T\ll T_c$), the uniaxial $g$-anisotropy provides only a finite $|c_2| \propto H$, with all higher order coefficients being zero. This occurs because the magnetic anisotropy arises only through the angular dependent effective field $\mathbf{H}_{\rm eff} = \mathbf{H}\cdot \mathbb{G}$, while the underlying susceptibility tensor $\partial \sum_i \langle S_i^\mu\rangle / \partial H_{\text{eff}}^\nu$ is constant and isotropic. 
 In the high-field polarized Region III, all coefficients $c_{2n}$ are finite and field-independent, indicating constant $\tau(\theta)/H$. This occurs because the polarized ground state is independent of field-strength, having spins aligned in the direction of $\mathbf{H}_{\text{eff}}$, i.e. $\langle \mathbf{S}_i\rangle = (1/2)\mathbf{H}_{\text{eff}}/|\mathbf{H}_{\text{eff}}|$. As a result, the high-field torque is given by $\tau \propto \text{d}|\mathbf{H_\text{eff}}|/\text{d}\theta$, which maintains a saw-tooth shape, independent of field strength. Finally, in the Curie limit ($T \gg T_c$), high temperature expansion indicates that contributions from $g$-anisotropy are suppressed as $|c_{2n}| \propto (H/T)^{2n-1}$. 

Lastly, we consider the case of $\Gamma_1>0$ but no $g$-anisotropy (taking $g_a = g_{c^*}$, $\Gamma_1 = +2.5$, and $K_1 = -5$). As shown via ED calculations in Fig.~\ref{fig:torque_models}(c), this model also displays a sharp saw-tooth torque in the vicinity of $H = H_c(90^\circ)$. Results of classical simulations to extract $c_{2n}$ are shown in Fig.~\ref{fig:torque_scale}(b). In Region I, all coefficients are finite and increase with field as $|c_{2n}| \propto H^{2n-1}$. However, since $c_2 \gg c_4 \gg c_6, ...$ the torque curves nearly follow a $\sin 2\theta$ form.
At high field (Region III), we find that all coefficients decrease with field as $|c_{2n}| \propto H^{-n}$, which leads to a relative suppression of the saw-tooth appearance with increasing field. This latter effect can be understood by considering the $H\to \infty$ limit, in which all spins are polarized along the field direction, i.e. $\langle \mathbf{S}_i \rangle = (1/2) \mathbf{H}/|\mathbf{H}|$, so that $\tau \propto \text{d}\left(\sum_{ij} H^\alpha H^\beta/|\mathbf{H}|^2\right)/\text{d}\theta\propto \sin 2\theta$. In this limit, $|c_4/c_2| \to 0$ for pure exchange anisotropy. Finally, in the high temperature paramagnetic regime, we find that $\Gamma_1$ exchange provides $|c_{2n}| \propto T^{-n}(H/T)^{2n-1}$. 

An important consequence of the above findings is that either $g$-anisotropy or bilinear exchange anisotropy is sufficient to induce sharp features in $\tau(\theta)$, as each guarantees an anisotropic $H_c(\theta)$. However, $g$-anisotropy tends to dominate at large $H$ and $T$ in terms of both the magnitude of $\tau/H$, and the relative deviations from $\sin 2\theta$ dependence. In particular, when both exchange and $g$-anisotropy are present, the asymptotic high-field, low-$T$ behaviour is $\lim_{H\to \infty}c_4/c_2 \sim C + \mathcal{O}(H^{-1})$, where the constant $C$ is a function of $g_{c^*}/g_{ab}$ only, shown in Fig.~\ref{fig:torque_models}(d).  Fits of high-field torque measurements may therefore directly yield the magnitude of the $g$-anisotropy, as demonstrated in more detail in \cite{sup}.

As a possible alternative to the previous calculations with the model Hamiltonian Eq.~\eqref{eq:Hamil},
 we have also considered the role of higher order 3-spin terms that can arise when a finite magnetic flux penetrates the 2D RuCl$_3$ layers\cite{motrunich2005variational,riedl2018critical}. Microscopically, such terms can appear from an uncompensated Peierls' phase, which modifies the $d-d$ hopping integrals $t \to t \ \text{Exp}[i\int\mathbf{A}\cdot \text{d}\vec{\ell}]$. In systems with weak spin-orbit coupling (SOC), this effect gives rise to interactions at order $t^3/U^2$ that couple $\mathbf{H}$ to the scalar spin chirality $\langle \mathbf{S}_i \cdot ( \mathbf{S}_j \times \mathbf{S}_k) \rangle$. The latter has been proposed as a possible hidden Ising order parameter giving rise to the observed torque response at high field\cite{modic2018chiral}. In $j_{\rm eff} = 1/2$ systems, strong SOC may modify the resulting 3-spin interactions, leading to a more general form: $\mathcal{H}_{ijk} = (\mathbf{H}\cdot \hat{n}) \sum_{\mu,\nu,\lambda} L_{ijk}^{\mu\nu\lambda} S_i^\mu S_j^\nu S_k^\lambda$, in terms of the unit vector $\hat{n}$, parallel to the $c^*$ direction for $\alpha$-RuCl$_3$. Here, the $(\mathbf{H}\cdot \hat{n})$ term arises from the angular dependence of the flux penetrating the honeycomb layers\cite{modic2018chiral,riedl2018critical}, which vanishes for in-plane fields. In order to estimate the coupling constants $L_{ijk}^{\mu\nu\lambda}$, we employed the exact diagonalization methods described in Ref.~\onlinecite{winter2016challenges} with electronic parameters suitable for $\alpha$-RuCl$_3$. Using this approach, we make three observations: (i) the components $L_{ijk}^{\mu\nu\lambda}$  differ strongly from the scalar spin chirality form (i.e. $L_{ijk}^{xyz} \neq L_{ijk}^{xzy}$ and $L_{ijk}^{xxx} \neq 0$), so that the field does not couple simply to $\langle \mathbf{S}_i \cdot (\mathbf{S}_j \times \mathbf{S}_k)\rangle$. (ii) Due to the angular dependence of $(\mathbf{H}\cdot\hat{n})$, the 3-spin terms would provide large $k(\theta)$ anomalies for in-plane field directions ($\theta = 0^\circ$), instead of the experimentally observed $\theta = 90^\circ$. However, (iii) we also find the 3-spin terms to be small, with the largest being $L \sim 10^{-5}$ meV/T, which would be associated with torque contributions at least an order of magnitude smaller than those computed taking only the $g$-anisotropy and bilinear couplings. For this reason, the 3-spin terms are unlikely to play a significant role in $\alpha$-RuCl$_3$.


On the basis of the above observations, we conclude that the minimal model for $\alpha$-RuCl$_3$  Eq.~\eqref{eq:Hamil}, including $\Gamma_1>0$ and $g$-anisotropy $g_{ab} > g_{c^*}$ captures the significant aspects of the observed out-of-plane magnetic torque, including: (i) the $\sin 2\theta$ dependence of $\tau$ in the low-field antiferromagnetic zigzag phase, and (ii) the saw-tooth angular dependence in the high-field polarized phase. However, both aspects are relatively universal.  
Since the strongly anisotropic magnetic Hamiltonians featured in $4d^5$ and $5d^5$ materials naturally lead\cite{janssen2017magnetization} to anisotropy in $H_c$, we can expect the saw-tooth $\tau$ to appear in a variety of ``Kitaev-candidate'' materials at high fields, including the iridates A$_2$IrO$_3$ (A = Na, Li).
For these materials, the scaling of high-field
 torque measurements may directly distinguish between the bilinear anisotropic interactions and the $g$-anisotropy, which is important for
establishing the full spin model.  

\textit{Acknowledgements.}$-$ 
We acknowledge useful discussions with K. Modic, B. Normand and M. Lee. The manuscript also benefited from discussion with C. Varma.
This work was supported by the Deutsche Forschungsgemeinschaft (DFG) through project SFB/TRR49.

\bibliography{torque}

\clearpage



\section*{Supplementary material}

# Supplemental Information for:
# On the saw-tooth torque in anisotropic $j_{\text{eff}} = 1/2$ magnets: Application to $\alpha$-RuCl$_3$

Kira Riedl,[1,*] Ying Li,[1] Stephen M. Winter,[1] and Roser Valentí[1]

[1]*Institut für Theoretische Physik, Goethe-Universität Frankfurt, Max-von-Laue-Strasse 1, 60438 Frankfurt am Main, Germany*

## Spin-spin correlations

In order to further characterize the two phases proposed in the phase diagram shown in Fig. 1(c) of the main text, we calculated with exact diagonalization (ED) the corresponding spin-spin correlations for a number of field strengths. The results are shown in Fig. S1.

At zero field, the $C_3$ symmetric model leads to three degenerate ordering wavevectors for the zigzag ground state, corresponding to the three high-symmetry points Y, M and M', shown in Fig. S1(a). From ED calculations, the zigzag phase (indicated by the dark blue background in Fig. S1) is characterized by large static correlations at these wavevectors. Classically, the wavevector degeneracy is not broken for out-of-plane fields $\mathbf{H}||c^*$ ($\theta = 90°$). As a result, correlations $\langle \mathbf{S_k} \cdot \mathbf{S_{-k}} \rangle$ at all three wavevetors remain equal for $\theta = 90°$. As the field is rotated towards the monoclinic $b$-axis, the zigzag domain with $\mathbf{k}$ = Y is energetically selected, leading to a relative enhancement of such correlations $\theta \neq 0$. Finally, in the asymptotically polarized phase (indicated by the light blue background in Fig S1), the zigzag correlations are suppressed, in favour of static correlations at $\mathbf{k} = \Gamma$.

## Field-Dependence of Torque and Magnetotropic Coefficient

For completeness, we also present the field dependence of the torque and magnetotropic coefficient for the full model at various fixed angles $\theta$ as a function of field strength $H$ (Fig. S2). The results may be compared with experimental data in e.g. supplemental figure 2 of Ref. 1 and Fig. 4 of Ref. 2. At intermediate field, the magnitude of the torque $\tau/H$ shows an angular-dependent kink at $H_c(\theta)$ marking the zigzag to polarized transition. At high field, $\tau/H$ essentially saturates, but decreases slowly towards the $H \to \infty$ limit. These results are consistent with the experimental data and with previous ED studies shown in the appendix of Ref. 3. In contrast, the magnetotropic coefficient $k/H$ shows a stronger feature with a characteristic dip and jump at $H_c$. This behaviour is similar to the experimental response of Ref. 2. Given that the present model reproduces the main features of the out-of-plane torque, this experiment has not detected the recently proposed narrow intermediate phases lying between the zigzag and quasi-polarized states[4–6], as these phases are explicitly absent in the present model[7].

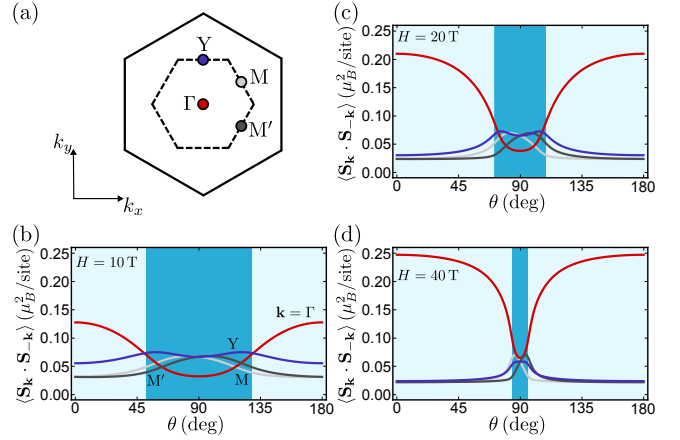

Fig. S1. (a) First (dashed) and second (solid) Brillouin zone of the honeycomb lattice with high-symmetry points $\Gamma$, Y, M, M' indicated. (b-d) Angle dependence of static spin-spin correlations $\langle \mathbf{S_k} \cdot \mathbf{S_{-k}} \rangle$ from ED calculations for high-symmetry points at (b) $H = 10\,\text{T}$, (c) $H = 20\,\text{T}$, and (d) $H = 40\,\text{T}$. The phase boundaries shown in Fig. 1(c) of the main text are indicated by the coloured background.

## Magnetic Torque in the Strong Anisotropy Limit

In order to understand the source of saw-tooth features in magnetic torque generally, it is useful to consider the behaviour of the torque in the limit where the field couples anisotropically to a single Ising variable $\langle \mathcal{O} \rangle = \pm 1$ representing the expectation value of the operator $\mathcal{O}$. There are many situations in which the energy is given by:

$$E = C\,(\mathbf{H} \cdot \hat{n})\,\langle \mathcal{O} \rangle \quad \text{(S1)}$$
$$= C|\mathbf{H}|\langle \mathcal{O} \rangle \cos(\theta) \quad \text{(S2)}$$

for some constant $C$, and $\theta$ is the angle between the magnetic field $\mathbf{H}$ and some unit vector $\hat{n}$. In this case, the energy is minimized for $\langle \mathcal{O} \rangle = -\operatorname{sgn}[C \cos(\theta)] = \pm 1$. That is, the Ising variable chooses the orientation that gives $E < 0$. The minimum energy and torque then follow:

$$E_{\min} = -C|\mathbf{H}|\operatorname{sgn}[C \cos(\theta)]\cos(\theta) \quad \text{(S3)}$$
$$\frac{\tau}{H} = \frac{1}{H}\frac{dE_{\min}}{d\theta} = -|C|\operatorname{sgn}[\cos(\theta)]\sin(\theta) \quad \text{(S4)}$$

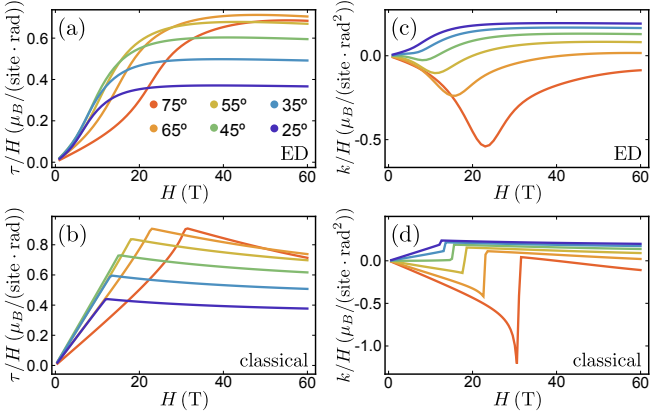

Fig. S2. Computed properties as a function of field of the model Eq. (2) of the main text for the parameters $(J_1, K_1, \Gamma_1, J_3) = (-0.5, -5.0, +2.5, +0.5)\,\text{meV}$ and $(g_a, g_b, g_{c^*}) = (2.3, 2.3, 1.3)$. Torque, normalized with field, for several angles $\theta$ within the $bc^*$-plane from (a) exact diagonalization and (b) classical calculations. Magnetotropic coefficient $k = \mathrm{d}\tau/\mathrm{d}\theta$, normalized with field, from (c) ED and (d) classical calculations.

The torque can be decomposed as:

$$\frac{\tau}{H} = -|C| \sum_{n=1}^{\infty} a_{2n} \sin(2n\theta) \tag{S5}$$

$$a_{2n} = \frac{4\cos(n\pi)\left[\sin(n\pi) - 2n\right]}{\pi(4n^2 - 1)} \tag{S6}$$

This form of the torque exhibits a significant saw-tooth appearance, displaying a jump in $\tau(\theta)$ at $\theta = 0$ due to the reversal of $\langle \mathcal{O} \rangle$ at this angle. This insight, by Modic et al.[8], led the authors to propose a hidden order in the high-field phase of $\gamma$-Li$_2$IrO$_3$ and $\alpha$-RuCl$_3$, identifying $\langle \mathcal{O} \rangle$ with the scalar spin chirality $\langle \mathbf{S}_i \cdot (\mathbf{S}_j \times \mathbf{S}_k) \rangle$. However, similar angle dependence may be derived for any Ising variable, so that $\langle \mathcal{O} \rangle$ may also represent a more conventional magnetic operator in the limit of strong bilinear exchange anisotropy, or $g$-anisotropy.

In order to demonstrate this point, we can consider the case of a single spin $S = 1/2$ with $g$-anisotropy in the presence of a magnetic field, described by the Hamiltonian:

$$\mathcal{H} = -\mathbf{H}^{\text{T}} \cdot \mathbb{G} \cdot \mathbf{S} \tag{S7}$$

$$\mathbb{G} = \begin{pmatrix} g_{xx} & 0 & 0 \\ 0 & g_{yy} & 0 \\ 0 & 0 & g_{zz} \end{pmatrix} \tag{S8}$$

$$\mathbf{H} = \begin{pmatrix} |\mathbf{H}|\sin\theta\cos\phi \\ |\mathbf{H}|\sin\theta\sin\phi \\ |\mathbf{H}|\cos\theta \end{pmatrix} \tag{S9}$$

where superscript T denotes the transpose. The energy is minimized for $\langle \mathbf{S} \rangle = (1/2)(\mathbb{G}^{\text{T}} \cdot \mathbf{H})/|\mathbb{G}^{\text{T}} \cdot \mathbf{H}|$, so that:

$$E_{\min} = -\frac{1}{2}\frac{\mathbf{H}^{\text{T}} \cdot \mathbb{G} \cdot \mathbb{G}^{\text{T}} \cdot \mathbf{H}}{|\mathbb{G}^{\text{T}} \cdot \mathbf{H}|} \tag{S10}$$

Specifying to the case of extreme $g$-anisotropy, (i.e. $g_{xx} = g_{yy} = 0$), then $\mathbf{H}^{\text{T}} \cdot \mathbb{G} = g_{zz}|\mathbf{H}|\cos\theta$. It is then clear that the minimum energy follows:

$$E_{\min} = -\frac{g_{zz}}{2}|\mathbf{H}|\,\text{sgn}\left[g_{zz}\cos\theta\right]\cos\theta \tag{S11}$$

which has precisely the form of Equation (S3). This result demonstrates that a sharp saw-tooth torque can arise even for an isolated spin. Although $\mathbf{S}$ is, in principle, a vector operator, the Hamiltonian with $g_{xx} = g_{yy} = 0$ permits only two ground states having $\langle S_z \rangle = \pm 1/2$.

A similar observation can also be made for exchange anisotropy. We can consider a pair of $S = 1/2$ spins with a ferromagnetic Ising interaction:

$$\mathcal{H} = -4J S_1^z S_2^z - \mathbf{H}^{\text{T}} \cdot (\mathbf{S}_1 + \mathbf{S}_2) \tag{S12}$$

In the limit $J \gg H$, the lowest lying states are $|\uparrow\uparrow\rangle$ and $|\downarrow\downarrow\rangle$, with energies $E_{\pm} \approx -J \pm |\mathbf{H}|\cos\theta$. Which of these states is the ground state depends on the orientation of $\mathbf{H}$, leading to a minimum energy:

$$E_{\min} \approx -J - |\mathbf{H}|\text{sgn}\left[\cos(\theta)\right]\cos(\theta) \tag{S13}$$

such a condition also leads to saw-tooth torque. In the opposition limit $H \gg J$, instead there is a unique ground state where the spins follow precisely the orientation of the field, giving $\langle S_1^z \rangle = \langle S_2^z \rangle = (1/2)\cos(\theta)$. As a result,

$$E_{\min} \approx -J\cos^2(\theta) - |\mathbf{H}| \tag{S14}$$

In this limit, the torque obeys a conventional $\sin(2\theta)$ dependence.

In order to further demonstrate this effect, we computed the classical torque for a series of hypothetical models exhibiting a ferromagnetic ground state at all fields. We emphasize that these models are not directly relevant to the real materials at low field, but are nonetheless instructive to consider in the context of the high-field polarized phases. In particular, we considered a 2D honeycomb model and employed $J_1 = -1, K_1 = 0, J_3 = 0$, while varying separately the effects of $g$-anisotropy and the $\Gamma_1$ term.

For the first series of calculations, shown in Fig. S3(a), we took $\Gamma_1 = 0$, $g_{ab} = 2$, and varied the value of $g_{c^*}$ in the range 0 to 2, while keeping $H$ constant. In this case, there is no exchange anisotropy, and the torque arises only from the anisotropic $g$-tensor. The angle dependence of $\tau(\theta)$ is the same for all field strengths. As in the main text, we rotated the field between the $b$-axis and $c^*$-axis. In the extreme limit $g_{c^*} \to 0$, the field only couples to the in-plane component of the uniform magnetization $\mathbf{m} \equiv 2^N \sum_i^N \langle \mathbf{S}_i \rangle$. For field orientations with $\mathbf{H} \cdot \hat{b} > 0$, the

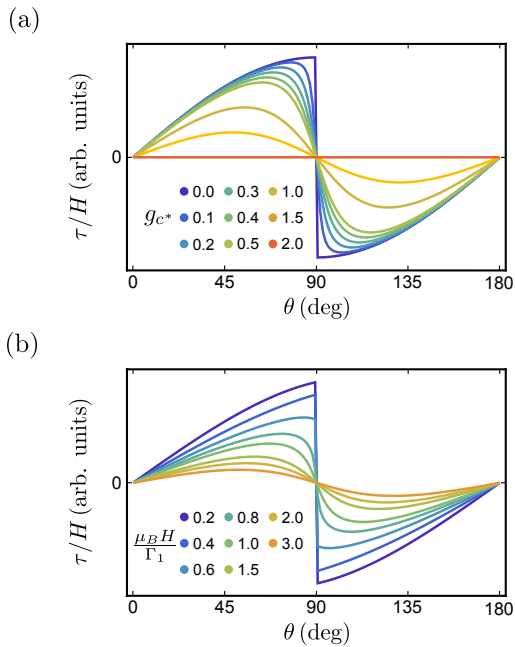

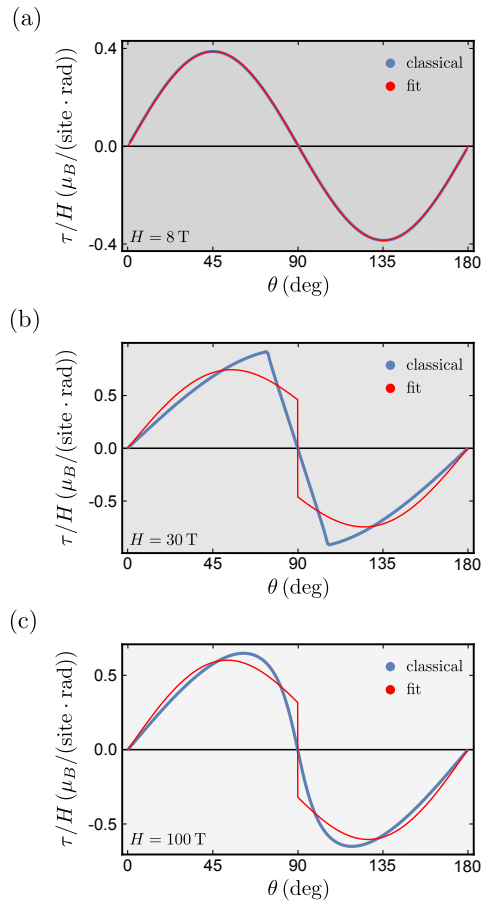

Fig. S3. Theoretical torque curves for the hypothetical models discussed in the text, demonstrating the effects of (a) relative $g$-anisotropy ($g_{ab} = 2$) and (b) exchange anisotropy in the ferromagnetic phase.

Fig. S4. Results of fitting the phenomenological expression of Ref. 8 to the classical torque curves obtained for the model of $\alpha$-RuCl$_3$ discussed in the main text, for different field regions: (a) Region I: $H < H_c(0°)$, (b) Region II: $H_c(0°) < H < H_c(90°)$, and (c) Region III: $H > H_c(90°)$.

energy is minimized for $\mathbf{m} = +\hat{b}$. In contrast, for $\mathbf{H} \cdot \hat{b} < 0$, the energy is minimized for $\mathbf{m} = -\hat{b}$. As a result, $\langle \mathbf{m} \cdot \hat{b} \rangle = \pm 1$ serves as an effective Ising variable in the limit of strong $g$-anisotropy for this field rotation. As $g_{c^*} \to 0$, the torque approaches the extreme saw-tooth form of Eq. (S4). This demonstrates that $g$-anisotropy is, in principle, sufficient to provide a saw-tooth torque in the high-field polarized phase.

For the second series of calculations (Fig. S3(b)), we employed $g_{ab} = g_{c^*} = 2$, and $\Gamma_1 = 0.1 \ll |J_1|$ in order to ensure a ferromagnetic ground state at zero field. We then varied the field strength in the range $\mu_B H/\Gamma_1 = 0.2$ to 3.0. In the limit $\mu_B |\mathbf{H}| \gg \Gamma_1$, the magnetization $\mathbf{m}$ follows the field precisely. As a result, the torque exhibits only a conventional $\sin 2\theta$ component due to the easy $ab$-plane anisotropy enforced by the finite $\Gamma_1 > 0$. For the opposite limit $\mu_B |\mathbf{H}| \ll \Gamma_1$, a different behaviour is observed. In this case, the magnetization is confined to the $ab$-plane, and $\langle \mathbf{m} \cdot \hat{b} \rangle = \pm 1$ again serves as an effective Ising variable, giving rise to a sharp saw-tooth feature. This demonstrates that exchange anisotropy may also give rise to an extreme saw-tooth torque provided the field remains small compared to the magnitude of the easy-plane or easy-axis anisotropy.

### Fitting Functions for Magnetic Torque

In Ref. 8, the authors extracted the saw-tooth component of the torque by fitting to the expression:

$$\frac{\tau}{H} = A_1 \,\mathrm{sgn}[\cos(\theta)] \sin(\theta) + A_2 \sin(2\theta) \qquad \text{(S15)}$$

where the first term represents the strong anisotropy limit, and the second term represents the weak anisotropy limit. In Fig. S4, we show the results of fitting this expression to the classical torque curves computed for the model for $\alpha$-RuCl$_3$ in the text, namely $J_1 = -0.5, K_1 = -5.0, \Gamma_1 = +2.5, J_3 = +0.5$, and $g_{ab} = 2.3, g_{c^*} = 1.3$. In the low field zigzag Region I for $H < H_c(0°)$, this expression yields reasonable fits, due to the fact that the torque nearly follows a conventional $\sin 2\theta$ curve. For the intermediate Region II, $H_c(0°) < H < H_c(90°)$, the fit function fails to reproduce the kinks at the phase boundaries, and therefore yields a somewhat poorer fit. Finally, for Region III, $H > H_c(90°)$, the fits are also somewhat

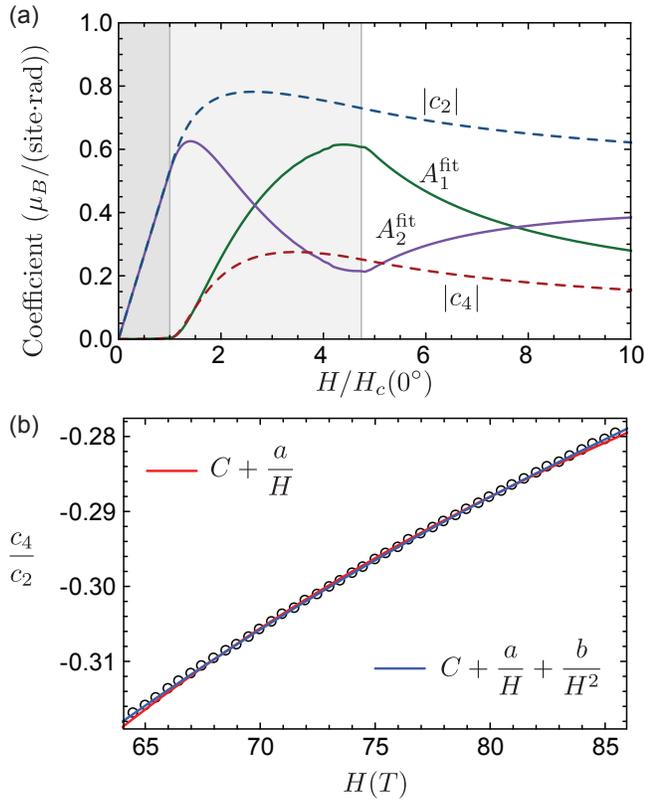

Fig. S5. (a) Field-dependence of coefficients obtained from fitting the classical torque curves obtained for the model of $\alpha$-RuCl$_3$ discussed in the main text. (b) Fitting of the high-field $c_4/c_2$ ratio, in order to demonstrate how to extract $g_{c^*}/g_{ab}$ from experimental torque data.

poor.

As discussed in the main text, the torque is more generally expanded as:

$$\frac{\tau}{H} = \sum_{n=1}^{\infty} c_{2n} \sin(2n\theta) \qquad (S16)$$

with the coefficients $c_{2n}$ obeying particular scaling with respect to $H$ and $T$. Least squares fitting of arbitrary torque curves by expression (S15) yields:

$$A_1^{\text{fit}} = \frac{\sum_{n=2}^{\infty} a_{2n} c_{2n}}{\sum_{n=2}^{\infty} a_{2n}^2} \quad , \quad A_2^{\text{fit}} = c_2 - a_2 A_1^{\text{fit}} \qquad (S17)$$

where the coefficients $a_{2n}$ are defined in equation (S6). Generally, such fit parameters do not obey any particular scaling. However, in the low-field Region I, where $c_2 \gg c_4 \gg c_6$, etc. the fitted coefficients will scale as: $A_2^{\text{fit}} \approx c_2$, and $A_1^{\text{fit}} \approx -1.215\, c_4$.

The field-dependence of $A_1, A_2, c_2$, and $c_4$ for the model for $\alpha$-RuCl$_3$ are shown in Fig. S5, as obtained from classical simulations. For $H < H_c(0°)$ (Region I), the magnitude of the $\sin 2\theta$ component grows as $A_2^{\text{fit}} \propto H$, while the saw-tooth component grows as $A_1^{\text{fit}} \propto H^3$, in accordance with the scaling of $c_2$ and $c_4$, respectively. In the intermediate field Region II where polarized and zigzag states both exist ($H_c(0°) < H < H_c(90°)$), the apparent saw-tooth component $A_1$ grows substantially, peaking at $H_c(90°)$. Finally, in the high-field Region III ($H > H_c(90°)$), the evolution of the $A_1/A_2$ ratio depends on the relative strength of different contributions. For pure $g$-anisotropy, this ratio is field-independent, leading to an extended field range where the saw-tooth shape remains. For pure exchange anisotropy, the ratio $A_1/A_2$ slowly decreases, as the $\sin 2\theta$ dependence is restored in the polarized phase for the $H \to \infty$ limit. A somewhat similar behaviour was observed in Ref. 8 for $\gamma$-Li$_2$IrO$_3$, suggesting a possible similar origin.

Finally, we demonstrate the experimental extraction of the $g$-anisotropy from high-field torque data, as discussed in the main text. Fig. S5 shows the dependence of the ratio $c_4/c_2$ obtained from classical simulations in Region III for experimentally accessible field strengths. Fitting this ratio to a polynomial of $(1/H)$ yields limiting values $\lim_{H \to \infty} c_4/c_2 = $ -0.12 to -0.16. From the theoretical expression for this ratio, plotted in Fig. 3(d) of the main text, we find $g_{c^*}/g_{ab} = 0.5$ to $0.6$. Finally, under the constraint $(2/3)g_{ab} + (1/3)g_{c^*} = 2$, we would estimate from this data that $g_{ab} = 2.3$ to $2.4$, and $g_{c^*} = 1.2$ to $1.4$. Since these values match the model employed in the simulations, it is demonstrated that the $g$-anisotropy may be extracted from high-field torque measurements, within experimentally accessible fields.



\end{document}